# Optimizing Generative AI's Accuracy and Transparency in Inductive Thematic Analysis: A Human-AI Comparison

**Matthew Nyaaba**[†1,2], **Min SungEun**[†3], **Mary Abiswin Apam**[†4], **Kwame Owoahene Acheampong**[5], **Emmanuel Dwamena**[6], **Xiaoming Zhai**[1, 7]

[1]AI4STEM Education Center, University of Georgia, Athens, GA, USA

[2]Department of Educational Theory and Practice, University of Georgia, Athens, GA, USA

[3]Department of Elementary, Middle Level, Library and Technologies Education, Kutztown University, Kutztown, PA, USA.

[4]Department; Midwifery Department, College of Nursing, Midwifery and Allied Health Sciences, Nalerigu, Ghana.

[5]Department of Lifelong Education, Administration, and Policy, University of Georgia, Athens, GA, USA

[6]Department of Curriculum and Instruction, University of Connecticut, USA

[7]Department of Science, Social Studies and Mathematics, University of Georgia, Athens, GA, USA

## Abstract

Thematic analysis is increasingly being conducted using Generative AI (GenAI), but challenges such as hallucinations, inaccuracies, and inconsistencies in various AI-driven approaches can undermine transparency and validation. To address these challenges, we employ the GPT-4 Turbo API integrated within a Stepwise Prompt-based Python Script to analyze interview data. This method is then compared with human coders' analysis. To carry out this comparison effectively, we followed a five-phase analysis and evaluation procedure. First, two independent human coders analyzed the interview transcript using the thematic analysis approaches of Clarke and Braun (2016) and Naeem et al. (2023). They followed a structured process of coding, theming, and interpreting the data. Next, an independent reviewer assessed the percentage similarity and differences between the two human coders' analyses to evaluate their consistency. For the third phase, GenAI was used to code, theme, and interpret the same interview data. Then, an independent reviewer compared the analysis generated by GenAI to the analysis from the human coders, rating the percentage of similarity and differences between the two. Finally, we conducted a comprehensive review of the entire process to address any errors or inconsistencies that were identified along the way. The results indicated that GenAI was highly effective in generating codes, categorizing themes, and providing broader interpretations, almost matching the performance of the human coders (with nearly 100% agreement in coding and theming). However, in the interpretation phase, while both GenAI and humans conducted a theme-based interpretation, GenAI took a more generalized and conceptual approach, whereas the human coders interpretations were more specific. Additionally, when both the human and GenAI processes were mapped to Naeem et al.'s (2023) six-step thematic analysis framework, it was found that GenAI followed four of the six steps, keywording, coding, and interpretation, while human coders followed the first three of these steps. These findings suggest that GenAI can be a highly viable tool for conducting inductive thematic analysis with minimal human intervention, showing promise in replicating human-driven processes with a high degree of accuracy.

*Keywords: Inductive Thematic Analysis, Qualitative Research, Transparency and Accuracy, Human-AI Comparison, Interpretation and Theming, AI.*

**Study Highlights**

*What is Already Known*

- Generative AI (GenAI) has demonstrated potential in inductive thematic analysis.
- A primary challenge in GenAI's inductive thematic analysis has been hallucinations and difficulty with interpretative nuances, which affect validation and transparency.
- Previous studies have explored OpenAI's API (GPT-3.5 Turbo) for thematic analysis. While this reduced hallucinations, token limits restricted the inclusion of supporting quotes, thus affecting transparency and validation.

*What This Study Offers*

1. Utilizes OpenAI's API (GPT-4 Turbo) to perform thematic analysis, reducing hallucinations.
2. Introduces a stepwise prompting strategy that systematically generates codes, themes, and interpretations to enhance transparency.
3. Conducts a rigorous evaluation process comparing human and GenAI analysis to provide comprehensive results.
4. Presents a clear methodological framework for using OpenAI API and stepwise prompts in inductive thematic analysis.

*What This Study Adds to the Field*

- Enhances transparency and accuracy in GenAI-driven qualitative research by ensuring traceability in coding and referencing.
- Develops a refined methodological approach, demonstrating the potential of stepwise prompt engineering for optimizing qualitative analysis.

## Introduction

Research shows that most qualitative studies use Generative AI (GenAI) such as ChatGPT for thematic analysis; however, they are prone to hallucinations and lack of transparency (Perkins & Roe, 2024; Anderson et al., 2024). There are studies that have employed OpenAI's API, GPT-3.5-Turbo, for more controlled and focused analysis. While this has been effective in reducing hallucinations, certain challenges still exist, such as token limits and the need for complex iterative prompt strategies to enhance transparency and validity (De Paoli, 2024). Qualitative research remains critical for understanding complex social phenomena, and its methodologies are evolving with the advent of advanced technological tools (Petre & Costa, 2024; Liu et al., 2024). Recent advancements have integrated artificial intelligence (AI) and large language models (LLMs) into computer-assisted qualitative data analysis software (CAQDAS), significantly improving the efficiency and depth of data analysis (Bryda & Costa, 2024).

However, the emerging applications of Generative AI (GenAI), such as ChatGPT, illustrate the potential to analyze qualitative data independently, without the need to be integrated within traditional CAQDAS platforms (Perkins & Roe, 2024). Despite the potential for AI to revolutionize qualitative analysis, studies indicate that GenAI presents both opportunities and challenges (Nyaaba et al. 2024). For example, Liu et al. (2024) explored zero-shot, few-shot, and context-driven prompt strategies using GenAI for coding virtual tutoring session transcripts. Their findings demonstrated variability in effectiveness, where specific prompting decisions notably impacted the quality of the generated outputs. Similarly, Perkins and Roe (2024) acknowledged that while GenAI thematic analysis can offer a richer and more nuanced understanding of themes, it still exhibits inconsistencies and hallucinations in its outputs, underscoring the limitations of the technology in its current state. Further evidence suggests that GenAI-driven coding may overlook nuanced, context-specific insights, reinforcing the need for rigorous human oversight and strategic prompt engineering (Zambrano et al., 2023; Liu et al., 2024). This highlights the ongoing challenges in ensuring accuracy, transparency, and contextual relevance when using GenAI for qualitative research.

To improve accuracy and transparency, this study employed GPT-4 Turbo, with the highest token limits available, and applied a stepwise, task-oriented prompting strategy. This approach ensured that GenAI-generated codes were supported by traceable statements and page references from the original transcript, increasing the validity and transparency of the analysis process. In contrast to other studies, this study utilized an original dataset, which had not been previously analyzed or made publicly available. This ensured that the thematic analysis was conducted on novel data and provided a comparative analysis with the human coders. Further, the study aims to contribute to the growing body of research by addressing the limitations commonly associated with chatbot-based GenAI and token limits in GenAI for qualitative research. It also seeks to overcome the challenges associated with identifying suitable prompt strategies, with the goal of ensuring that rigorous, transparent, and reliable thematic analysis can be conducted solely by GenAI or with minimal human revisions that can be used by both experts and emerging qualitative researchers. Th following research questions guided the study: How does the use of GPT-4 Turbo with stepwise prompting affect the transparency and validity of inductive thematic analysis? How do the inductive thematic analysis outputs generated by GenAI compare to those produced by human coders in terms of accuracy and transparency?

## Review of Previous Studies

### Thematic Analysis

Thematic analysis is a widely used qualitative research method that helps researchers identify, analyze, and report patterns (themes) within qualitative data. It is flexible and accessible, making it valuable across fields like psychology, education, sociology, and healthcare (Braun & Clarke 2006). The process involves familiarizing oneself with the data, generating initial codes, and searching for themes, followed by reviewing, defining, and naming the themes. Over time, thematic analysis has evolved from a simple descriptive method to a more complex analytical approach, with its flexibility allowing application across various data types. Although Braun and Clarke's (2006) framework are a popular guide, thematic analysis has seen ongoing development such as those Naeem et al. (2023) who provide a novel six-step for conceptual model development in qualitative research.

Amidst this popularity, there's no universally agreed-upon workflow for thematic analysis, and scholars continue to debate best practices, particularly around the distinction between *inductive and deductive* approaches and the role of theory in coding. For example, in an inductive approach, themes emerge directly from the data, whereas in a deductive approach, themes are guided by pre-existing theories (Naeem et al., 2023). Additionally, there are disagreements about coding granularity, with some researchers advocating for detailed, micro-level coding, while others prefer broader, thematic coding (Nowell et al. 2017). These discrepancies, along with issues like variability in theme identification, subjectivity in interpretation, and lack of transparency, have highlighted the need for more consistent, transparent, and reliable methods, areas where GenAI can make a significant impact.

### GenAI and Thematic Analysis

There have been significant advancements in research exploring the potential of GenAI in both quantitative and qualitative data analysis (Zambrano et al., 2023; Perkins & Roe, 2024). Several sstudies have shown that GenAI has the potential to provide more consistent outputs than human coders, especially when handling large datasets. This consistency stems from the AI's ability to avoid researcher fatigue and biases, ensuring reliable results over time (De Paoli, 2024; Acheampong & Nyaaba, 2024). This section reviews key studies that examined the role of GenAI in qualitative analysis, highlighting both its strengths and limitations. Perkins and Roe (2024) investigated the use of ChatGPT (GPT-4) for inductive thematic analysis, comparing traditional human coding with AI-assisted analysis. Their findings indicate that while GenAI enhances data processing and theme identification, it still requires human expertise for nuanced interpretation. Notably, GenAI-generated themes were aligned with human-coded results, demonstrating their potential to expedite the thematic analysis process. Similarly, other studies showed that using GPT-3.5 Turbo API in inductive thematic analysis of semi-structured interviews can infer most human-identified themes (De Paoli, 2024). A comparison between GenAI and traditional manual coding showed that GenAI generated successfully identified basic themes (Morgan, 2023). This aligns with Lee et al. (2024) and Nyaaba et al. (2024), which depict GenAI's role in diverse research phases such as coding, theme generation, preprocessing quotes, literature review, and ideation.

**Challenges in Using GenAI for Coding**

Despite its efficiency in qualitative data analysis, GenAI presents several challenges that require careful consideration. One key issue is inconsistencies and hallucinations, where GenAI generates themes or interpretations that are not present in the original data, as noted by Perkins and Roe (2024) and De Paoli (2024). This raises concerns about validity and the need for human verification. Additionally, studies such as Morgan (2023) and Lee et al. (2024) highlight that GenAI effectively identifies basic themes but struggles with subtle, interpretative nuances, making it less reliable for capturing deeper meanings within qualitative data.

Another challenge is token processing limits, which requires researchers to split large datasets into smaller segments, a scenario which can disrupt thematic coherence (De Paoli, 2024). Moreover, structured prompt engineering is essential for obtaining accurate and meaningful outputs, as emphasized by Bijker et al. (2024). Without carefully designed stepwise prompts, GenAI may produce vague or misaligned codes, necessitating iterative refinements. Lastly, GenAI still requires human oversight, particularly for validating codes and refining themes to ensure contextual accuracy (Lee et al., 2024; Morgan, 2023). Furthermore, while GenAI accelerates the coding process, its dependence on human expertise for interpretation and validation underscores its role as a complementary tool rather than a full replacement for human qualitative analysis.

## Method

This study aligns with the goals of comparative analysis by examining the similarities and differences in GenAI and human coders in inductive thematic analysis of an interview transcript. The primary source of data we used for analysing the interview transcripts came from a semi-structured interview with a healthcare professional at (hidden for review purpose) about her migration experiences as a nurse and midwife. The ethical approval for the use of the data came through adult consent and permission of the interviewee through a (school —blinded). The data did not undergo any substantive analysis and was specifically collected for this methodological demonstration. Therefore, the focus of the study was not on the outcomes of the interview transcript but on demonstrating the thematic analysis procedures between GenAI and humans. We followed a five-phase approach in this analysis and evaluation process (See Figure 2).

### Human Coding Process

The *first phase* involved human analysis, where two independent coders conducted inductive coding based on the Clarke and Braun (2016) and more particularly Naeem et al. (2023) systematic thematic analysis process, the six-step for qualitative research (See Figure 1). The human coders were given a maximum of six weeks to complete this process. The systematic thematic analysis process by Naeem et al. (2023, p.4) consists of six steps: (1) Selection of quotations, where relevant excerpts from transcripts are identified; (2) Identification of keywords, extracting significant terms; (3) Coding, assigning meaningful codes to keywords; (4) Theme identification, grouping related codes into broader themes; (5) Conceptualization, interpreting themes to develop theoretical concepts; and (6) Development of a conceptual model, presenting the final framework through arrows and boxes to visualize relationships, leading to the final conceptual framework and naming the model.

This approach allowed the human coders to generate initial codes, themes and interpretations from the qualitative data. The coders did a close reading coding, identifying codes from the transcripts. They further created tables with codes and their supporting statements. They proceeded with these codes and further interpreted them based on the generated themes. Each of the coders was given a deadline to submit the results. Initial results were collected, and feedback was provided by two experts. The coders finetuned their results based on the feedback. Following this, the *second phase* involved an independent reviewer and expert who performed an interrater reliability analysis between the two results of the two coders. This step required the reviewer to identify similar codes, themes, and interpretations between the two human coders. The independence rater used a maximum of four weeks to establish the rating outcomes of the two results.

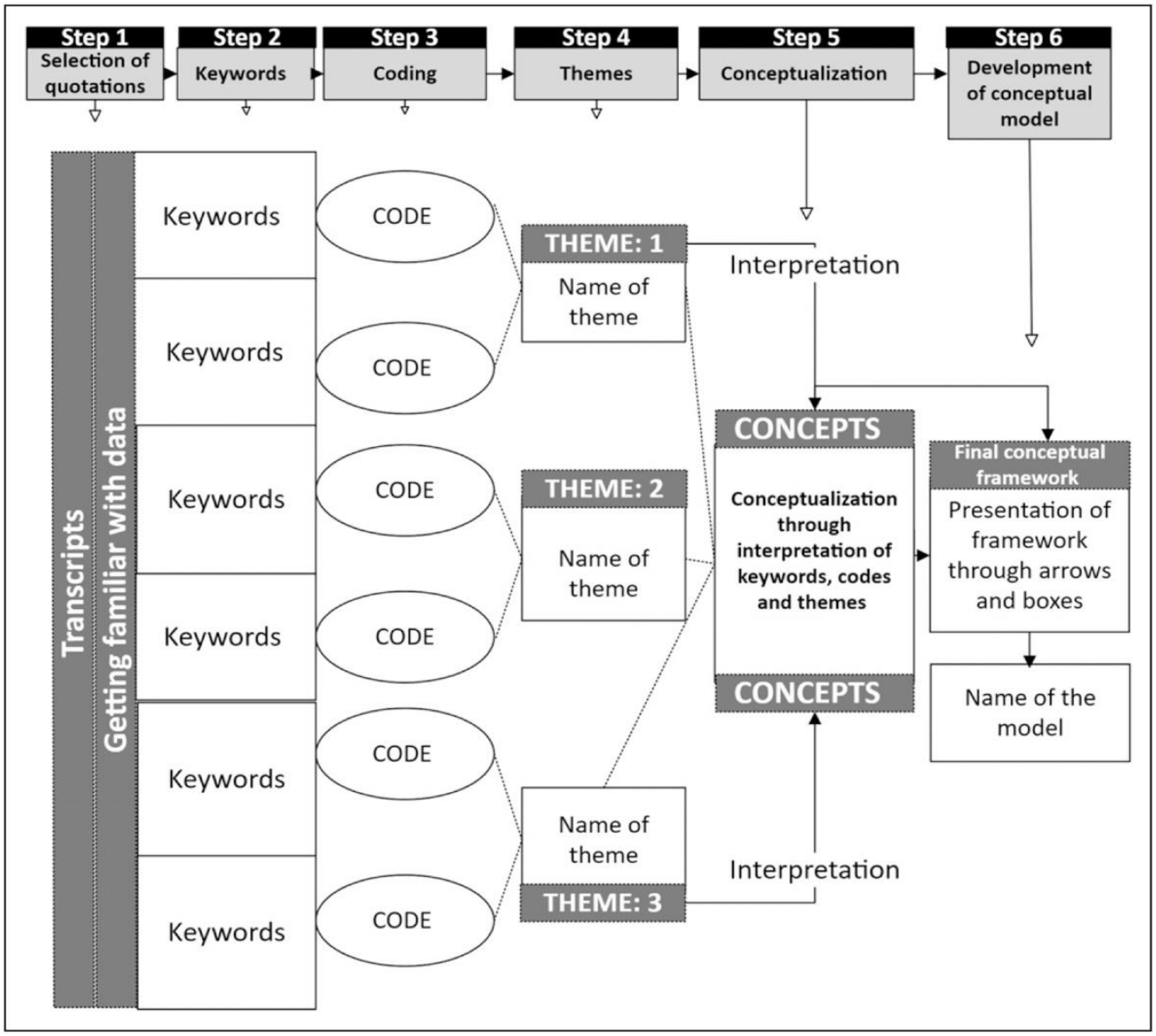

*Figure 1. A systematic thematic analysis process: A novel six-step for conceptual model development in qualitative research (Naeem et al., 2023, p.4)*

**Generative AI Coding Process**

The *third phase*, the GenAI coding process was conducted. A Python script was used to interact with the OpenAI GPT-4 turbo API, which processed the interview transcript, segmenting it into manageable chunks. The Python script is a file with a series of instructions written in Python programming language; it is executed by the Python interpreter to automate tasks, process data or algorithm. We run this prompt with the OpenAI API within the Jupyter environment. The use of the API aimed to reduce the inaccuracies and hallucinations present in the use of the ChatGPT chatbot. The script was embedded with *stepwise prompt* to extract relevant and emerging code based on the transcript focus. To ensure accuracy and traceability, GenAI provided supporting text and page references from the transcript. The *next phase*, the fourth, took 6 weeks and required another independent reviewer who performed the same inter-rater analysis by calculating the percentage similarity and differences between GenAI and the human coders. The *final phase*, the fifth, involved a 4-week review of the two previous reviews for accuracy in the inter-ratings of the two review processes.

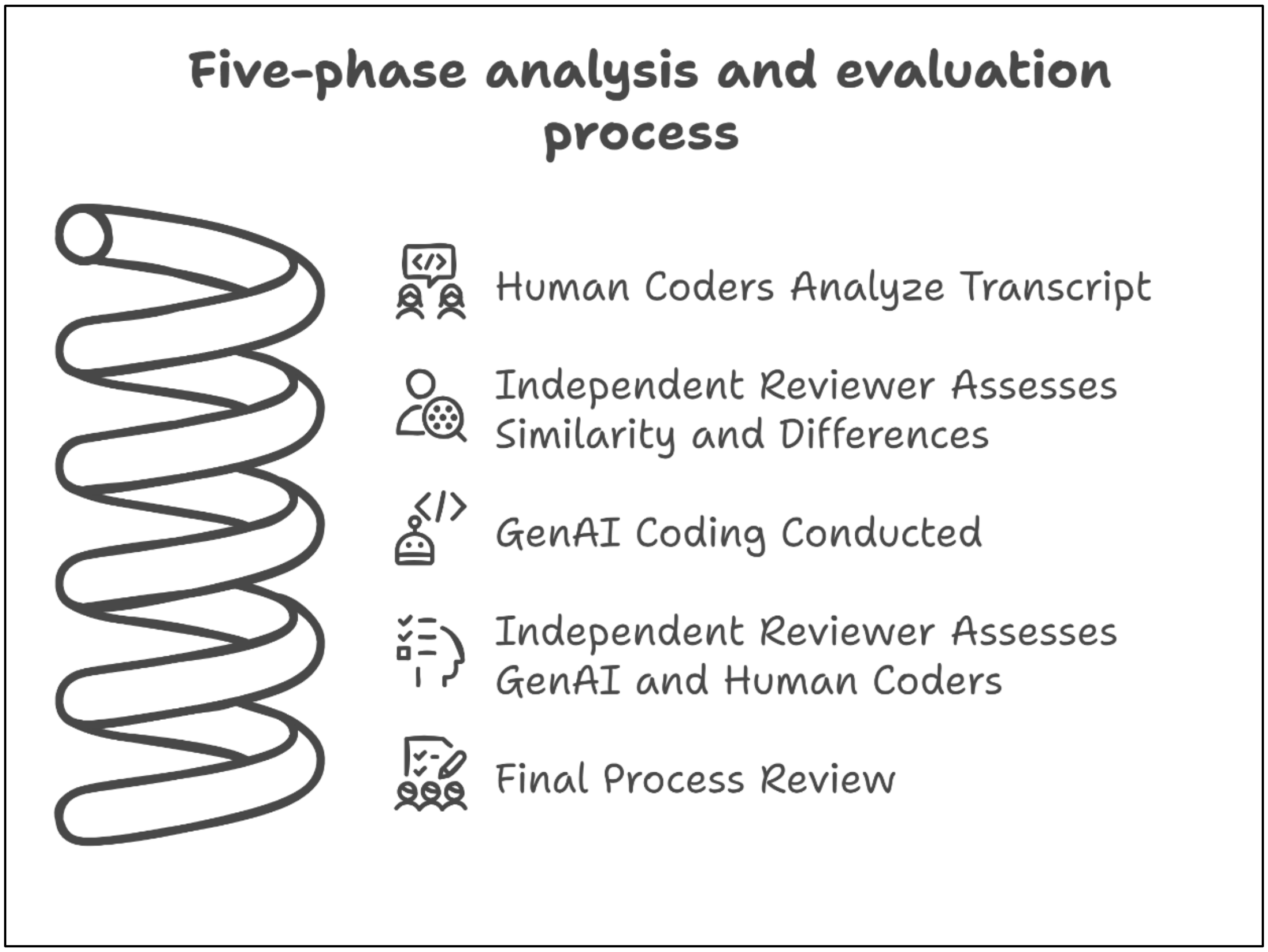

*Figure 2: Five Phase Analysis and Evaluation of Inductive Thematic Analysis, Comparison Between Human and GenAI*

*The GenAI and OpenAI*

GenAI can either be used by engaging with it on a chatbot interface like the ChatGPT or by engaging via the API. While GenAI chatbots excel in providing user-friendly, conversational experiences, making them better suited for simple, interactive tasks, using an API offers greater control, customization, and scalability for complex tasks. APIs enable more efficient automation and data processing, reducing the risk of "hallucinations" by ensuring predictable, accurate responses, especially when integrated with verified data sources (Perkins et al 2024; Liu et al, 2024).

*Data Preparation*

The data for this analysis consisted of semi-structured interviews, which were collected as part of a larger study on the migration experiences of healthcare professionals. These interviews were stored in a Microsoft Word document format (.docx). To facilitate the analysis, the document was divided into smaller segments for processing. Each "*page*" of data was defined as 10 paragraphs, ensuring manageable chunks of qualitative information that could be fed into the AI model for analysis. This preprocessing step ensures that the data is chunked into manageable segments, making it easier to pass specific text segments to the API for analysis. It simplifies the process of working with large datasets.

*Stepwise Prompt Strategy*

Stepwise prompt strategies can be linked to various analytical references in quantitative analysis, particularly in fields such as regression. For instance, stepwise regression is a statistical method used to select the most significant variables for inclusion in a predictive model (Draper, & Smith, 1998). Stepwise strategies can also be linked to process tracing in qualitative research, where events or processes are broken down into steps for analysis (Wohlrab et al. 2020; Cárdenas & Valcárcel, 2005). In this study stepwise prompt strategy is a method of structuring prompts in a sequential, organized manner to guide a model or AI through a series of steps to accomplish a task or generate a desired output. In this approach, the task is broken down into smaller, manageable stages or instructions, with each step building upon the previous one.

To guide GenAI in performing the thematic analysis, a *stepwise prompt* was crafted to generate the most accurate results of codes, themes, interpretation (See Figure 3). The process began with a *Python script* that was used to handle the qualitative interview data stored in a *Word document (docx)* format. This script broke the document down into manageable *"pages"*, with each page containing approximately 10 paragraphs. By segmenting the data in this way, the analysis was made more manageable, allowing the GenAI to focus on smaller chunks of text, which facilitated a more detailed and accurate extraction of themes.

The GenAI was then given specific instructions to extract only the most *relevant and inductively emerging codes* from the text. The prompt emphasized that the GenAI should capture distinct and meaningful ideas, patterns, or observations that were directly related to the *migration experiences* of nurses and midwives, particularly those migrating from developing to developed countries. By focusing on these central themes, the GenAI was guided to identify patterns that were not only significant but also aligned with the objectives of the research. Further, the prompt directed GenAI to avoid generating a code for every minor observation or less meaningful detail. Instead, the model was encouraged to emphasize *only the most prominent and relevant themes*, ensuring that the thematic analysis would capture the core ideas that were central to the research. This helped streamline the process by eliminating irrelevant or redundant codes, ensuring that the AI's output was focused on the most critical findings.

To enhance the *traceability* and v*erifiability* of the GenAI-generated codes, the prompt instructed the AI to return each emerging code along with *the exact sentence or passage* from which it was derived. In addition, the corresponding *page number* was included to ensure easy reference and validation. This structure allowed the themes generated by the AI to be cross-checked against the original data, providing a transparent and reliable way to verify the accuracy of the GenAI's thematic extraction. The traceability feature not only facilitated the validation of the GenAI's results but also ensured that the analysis remained grounded in the original qualitative data.

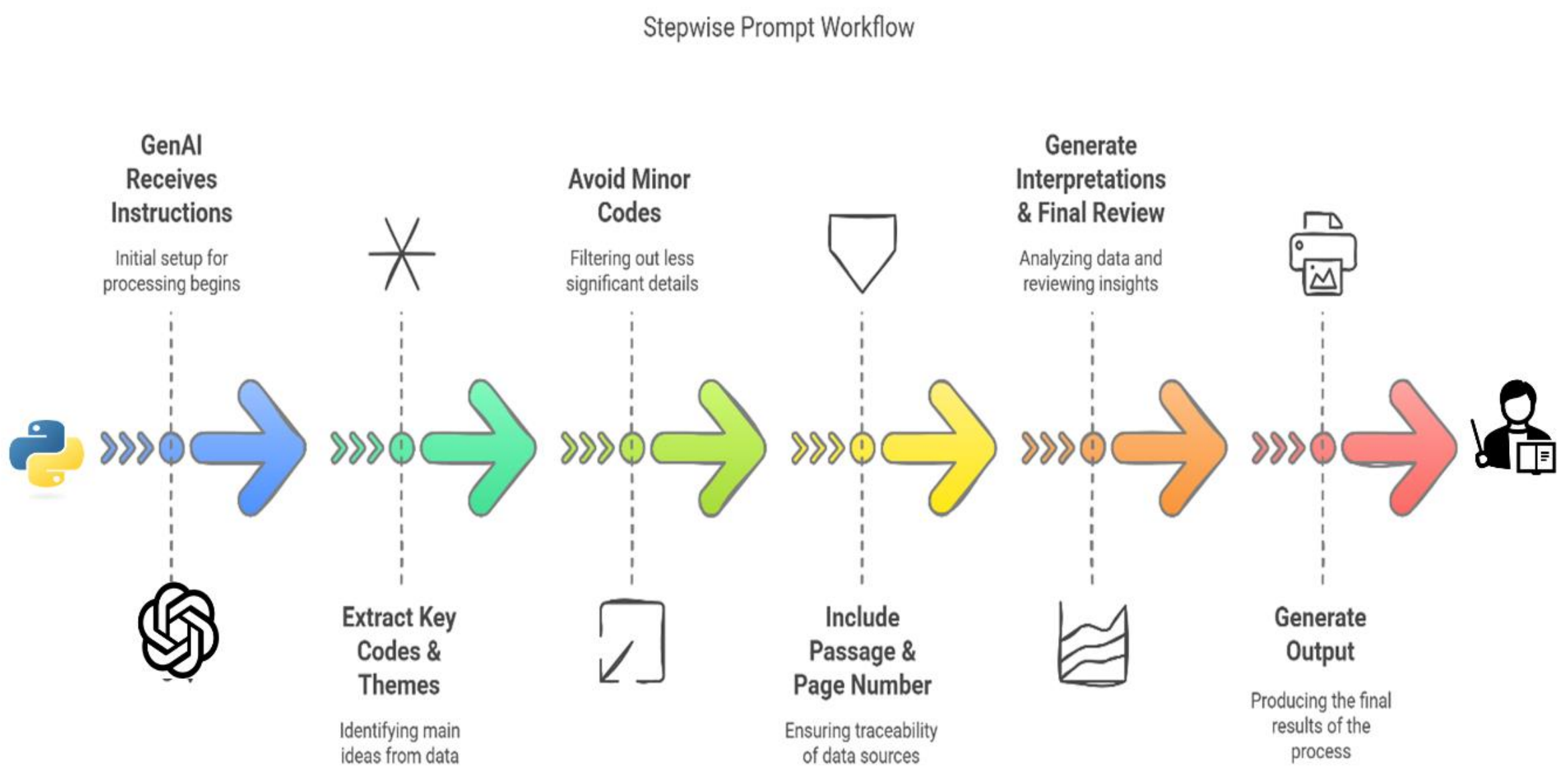

*Figure 3: Stepwise Prompt Workflow within the Python Environment*

*API Integration and Text Segmentation*

The OpenAI API (GPT-4 turbo) was used to perform thematic analysis through integration with Python scripts. We set the temperature at 0.3 to ensure focused and deterministic responses, minimizing randomness. Also, we configured it at a maximum token limit of 1000 to control the length of the generated output. We then assigned the role as *skilled qualitative researcher focused on inductively emerging codes*, guiding the model to extract meaningful and relevant codes from qualitative data (See Figure 4).

The process began by loading the .docx file containing qualitative data and splitting it into *10-paragraph segments*, treating each as a separate *page* for analysis. For each segment, the Python script instructed the API to analyze the text and generate thematic *codes* based on a specifically formulated prompt. GenAI then returned thematic codes along with the supporting text and corresponding *page numbers*, ensuring traceability between the generated themes and their source ensuring *contextual output*.

.

```python
import openai
from docx import Document

# Set your OpenAI API key
openai.api_key = 'sk-'  # Replace with your actual API key

def extract_emerging_codes(text_segment, page_number):
    prompt = (
        "Analyze the following qualitative data and extract only the most relevant, inductively emerging codes that capture "
        "distinct and meaningful ideas, patterns, or observations about the migration experiences of nurses and midwives from "
        "developing countries to developed countries. Avoid generating a code for every observation. For each emerging code, "
        "provide the code as a concise phrase or keyword, followed by the exact sentence or passage it was derived from and 'Page {page_number}' "
        "to ensure traceability.\n\n"
        f"Text Segment:\n{text_segment}\n\n"
        "Emerging Codes with Supporting Sentences and Page Number:"
    ).format(page_number=page_number)

    response = openai.ChatCompletion.create(
        model="gpt-4-turbo",
        messages=[
            {"role": "system", "content": "You are a skilled qualitative researcher focusing on inductively emerging codes."},
            {"role": "user", "content": prompt}
        ],
        max_tokens=1000,
        temperature=0.3
    )

    # Extract the codes with traceability information from the response
    codes_with_context = response['choices'][0]['message']['content'].strip()
    return codes_with_context

# Load the Word document using the specified file path
file_path = r"C:\Users\matth\OneDrive - University of Georgia\DOCUMENTS\Transcript.docx"
doc = Document(file_path)

# Split the document into "pages" (every 10 paragraphs as a new "page")
page_size = 10
pages = [doc.paragraphs[i:i + page_size] for i in range(0, len(doc.paragraphs), page_size)]

# Generate all emerging codes with supporting sentences and page numbers
all_codes_with_traceability = []

for page_number, page_content in enumerate(pages, start=1):
    text_segment = "\n".join([paragraph.text for paragraph in page_content if paragraph.text])
    codes_with_context = extract_emerging_codes(text_segment, page_number)
    all_codes_with_traceability.append(f"Page {page_number}:\n{codes_with_context}\n")

# Combine for Output
detailed_output = "\n".join(all_codes_with_traceability)
print("All Emerging Codes with Supporting Sentences and Page Numbers:\n", detailed_output)
```

*Figure 4: API Configuration for Qualitative Data Extraction Using GPT-4 Turbo*

**Analysis**

The results from the human coders and GenAI were compared, with a focus on inter-rater reliability to assess the percentage of similarity and differences of the codes. The themes and interpretation were rated based on consensus. In qualitative research, inter-rater reliability refers to the level of agreement between multiple researchers or coders when independently analyzing and categorizing qualitative data, such as interview transcripts or open-ended survey responses (Barbour, 2001; Cohen, 1960).

**Results**

## GenAI Coding

GenAI successfully captured many of the core ideas recognized by the human reviewers, which is particularly notable given the substantial length of the single transcript. The GenAI's output was compiled into a final document, which provided the full list of emerging codes along with their supporting sentences and page references. This structure facilitated the analysis of the themes in relation to the research questions, ensuring that the results were aligned with the migration context of the healthcare professionals being studied.

*Emerging Codes with Supporting Sentences*

The first step in the script involves identifying the emerging codes from the qualitative data. These are specific ideas, phrases, or observations related to transcript. A total of fifty-nine (n=59) codes are paired with supporting sentences and page numbers from the transcript (See Figure 4a). For example, the interview began with a clear *Confidentiality Assurance*, where the interviewer explained, "So this interview is purely for academic purposes, and so whatever you would say would just be within the academic space. Your name and your identity will not be revealed." This assurance, provided on *Page 2*, was followed by a request for *Permission for Participation*: "So, do I still have your permission to start with the interview?" The interviewer then clarified the focus of the discussion, stating, "As I spoke to you earlier on, today our interview is going to be about migration, I'll basically be asking for the reasons that informed your decision to go to the UK, the experiences that you've gotten so far, and basically the differences, or the similarities between the healthcare system that you were practicing here in Ghana and that of the UK." Moving to *Page 3*, the participant shared an *Accidental Career Discovery*, saying, "I changed to midwifery and I fell in love with it." Finally, the influence of early life experiences came to light with the *Influence of Childhood Experiences* emerging code, as the participant recalled, "I had the opportunity to see a traditional birth attendant doing a delivery for a lady." The output shows a list of 15 codes which could considered for themes before proceeding to generate the themes (See Figure 4b). Following this, we found the process as following Naeem et al. (2023) six processes where the first 59 codes could be considered as keyword and the 15 emerging codes as code.

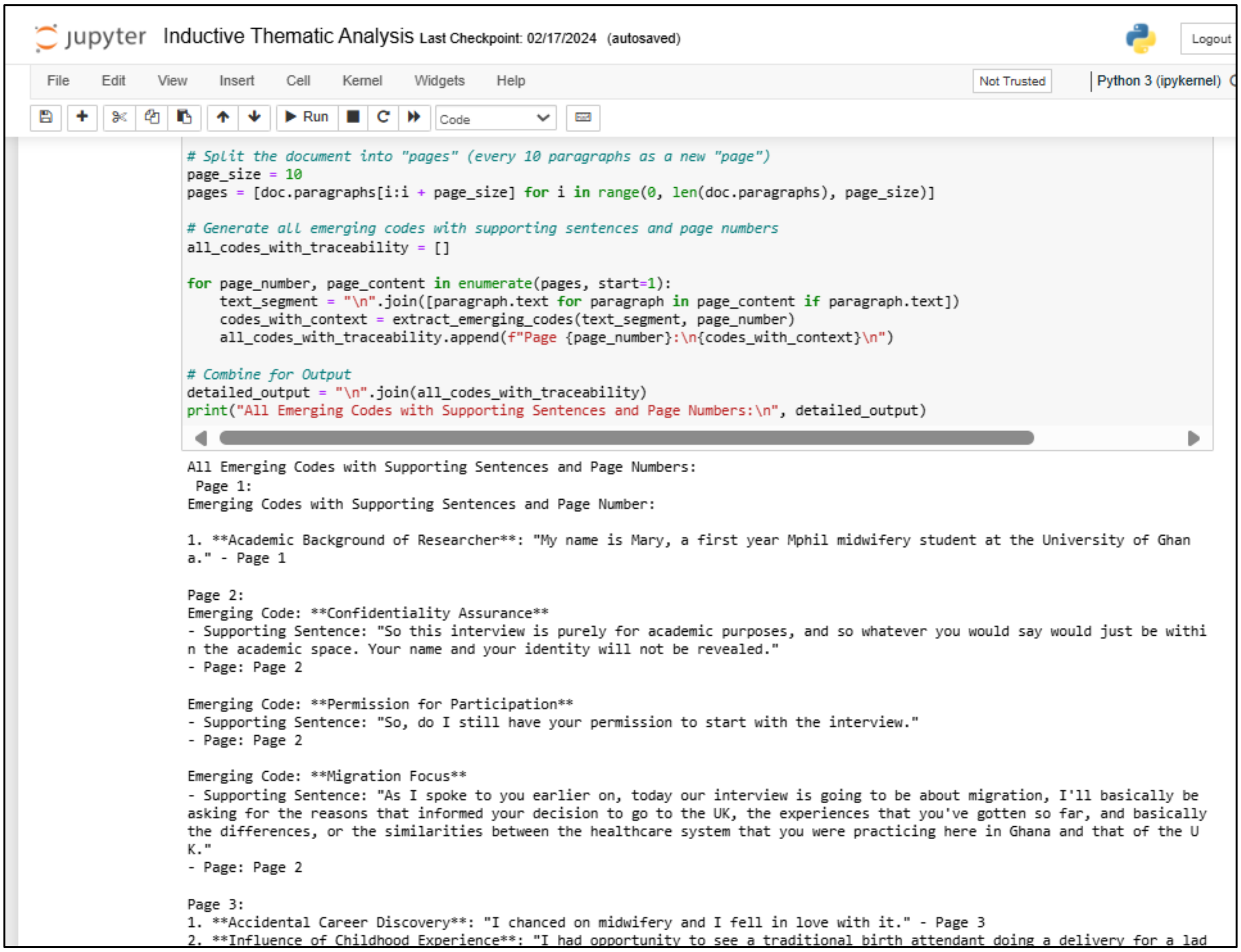

*Figure 4a: Emerging Codes with Supporting Sentences*

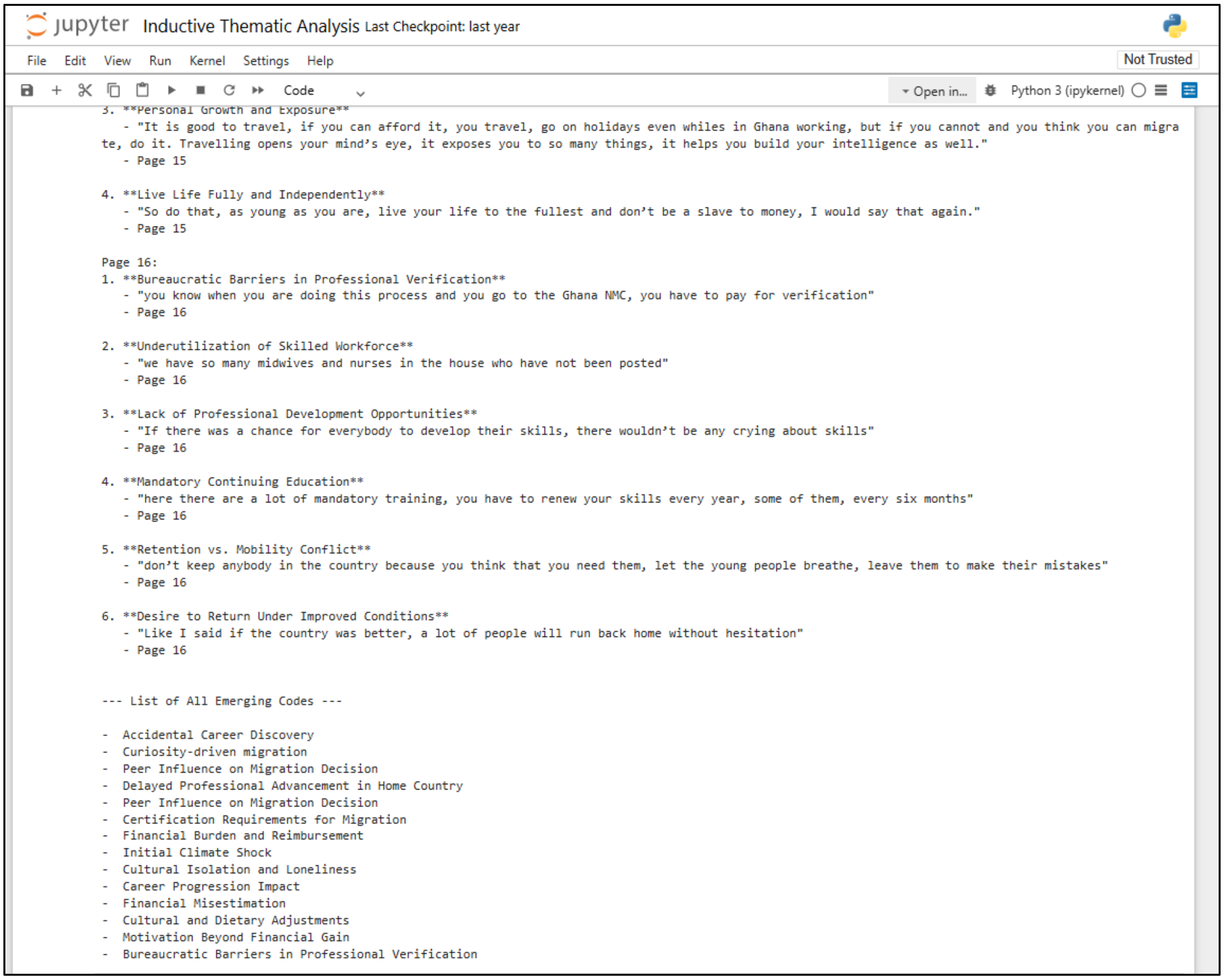

*Figure 4b: List of emerging Codes that can be Considered as Themes*

*GenAI Themes Generation*

After extracting the codes, the next step is to group them into *themes*. This output shows the *theme generation* process where the GenAI identifies four (4) overarching themes that capture the related codes (See Figure 5). These themes had varying numbers of codes associated with them. For example, the theme; *Personal and Professional Motivations for Migration was* comprised of a (n=7) codes. These codes, "curiosity-driven migration," "non-financial motivations," "peer influence on migration decision," "motivation through social support," and "motivation beyond financial gain," were all considered by the GenAI as codes associating with the theme. Also, from Figure 2b, the list of fifteen (15) emerging codes of GenAI that can be considered as themes included, *Accidental Career Discovery, Curiosity-driven migration, Peer influence on Migration Decision* etc.

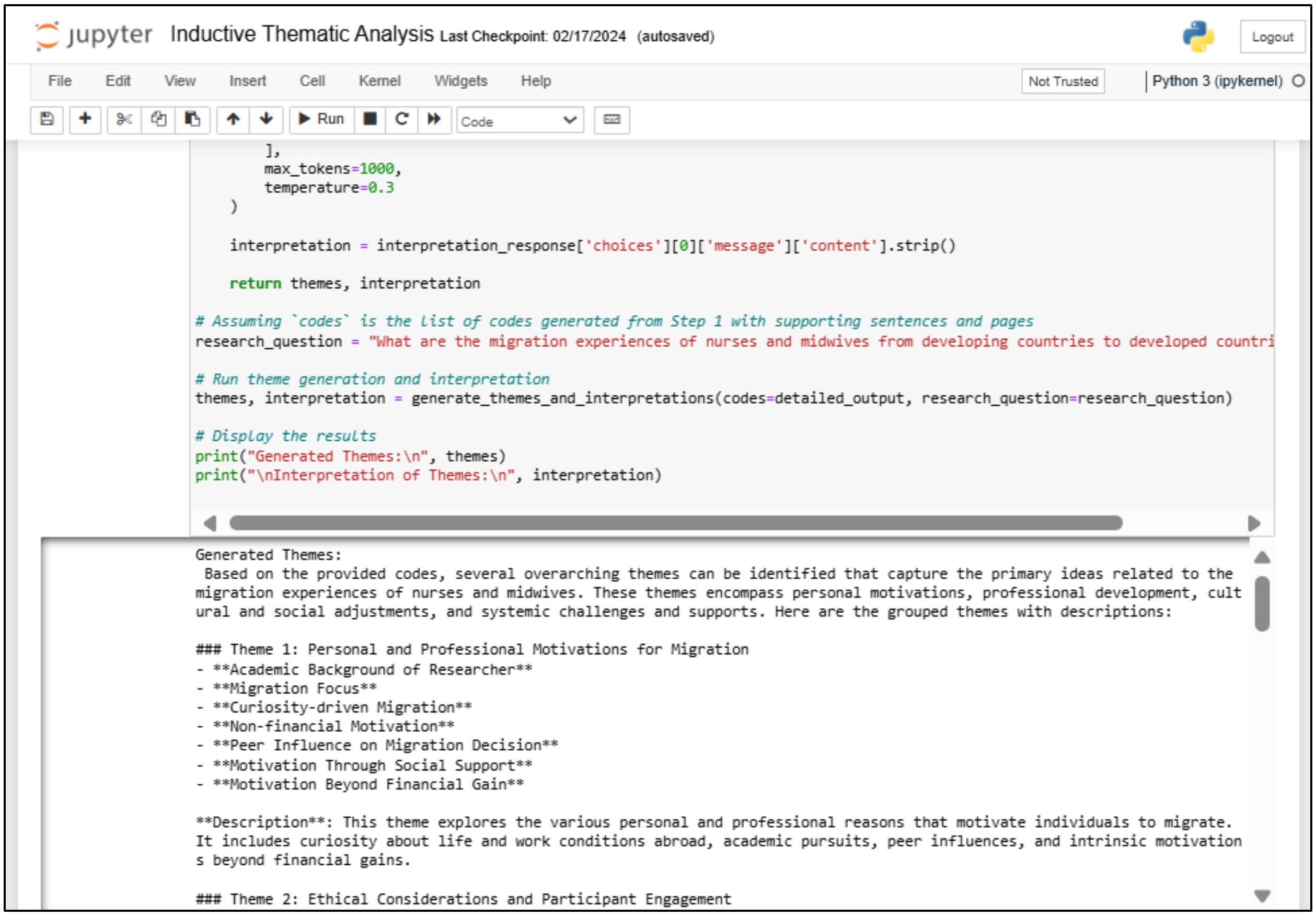

*Figure 5: Emerging Themes by GenAI*

*GenAI Interpretation*

The final step in the process is the interpretation of the generated themes. GenAI provides a detailed explanation of what each theme represents and how the codes contribute to the theme (See Figure 5). For example, interpretation of theme on *Personal and Professional Motivations for Migration*, this was what GenAI has to say: "

*Theme 1: Personal and Professional Motivations for Migration*

*This theme is central to understanding why nurses and midwives from developing countries choose to migrate to developed countries. The motivations are multifaceted, encompassing both personal and professional dimensions. Curiosity about life in different settings, influenced by academic pursuits or peer discussions, highlights a proactive approach to seeking new experiences and knowledge. This curiosity often extends beyond mere financial incentives, indicating a deeper desire for personal growth and development. The influence of peers and social support networks also plays a crucial role, as these individuals often rely on the advice and experiences of others who have migrated before them. This theme underscores the complex interplay of factors that drive*

*migration decisions, which are not solely based on economic benefits but also on professional enrichment and personal fulfillment*." _- (GenAI Interpretation)

This interpretation connects the thematic analysis to real-world experiences and provides a deeper understanding of the data. It helps to translate data into actionable insights.

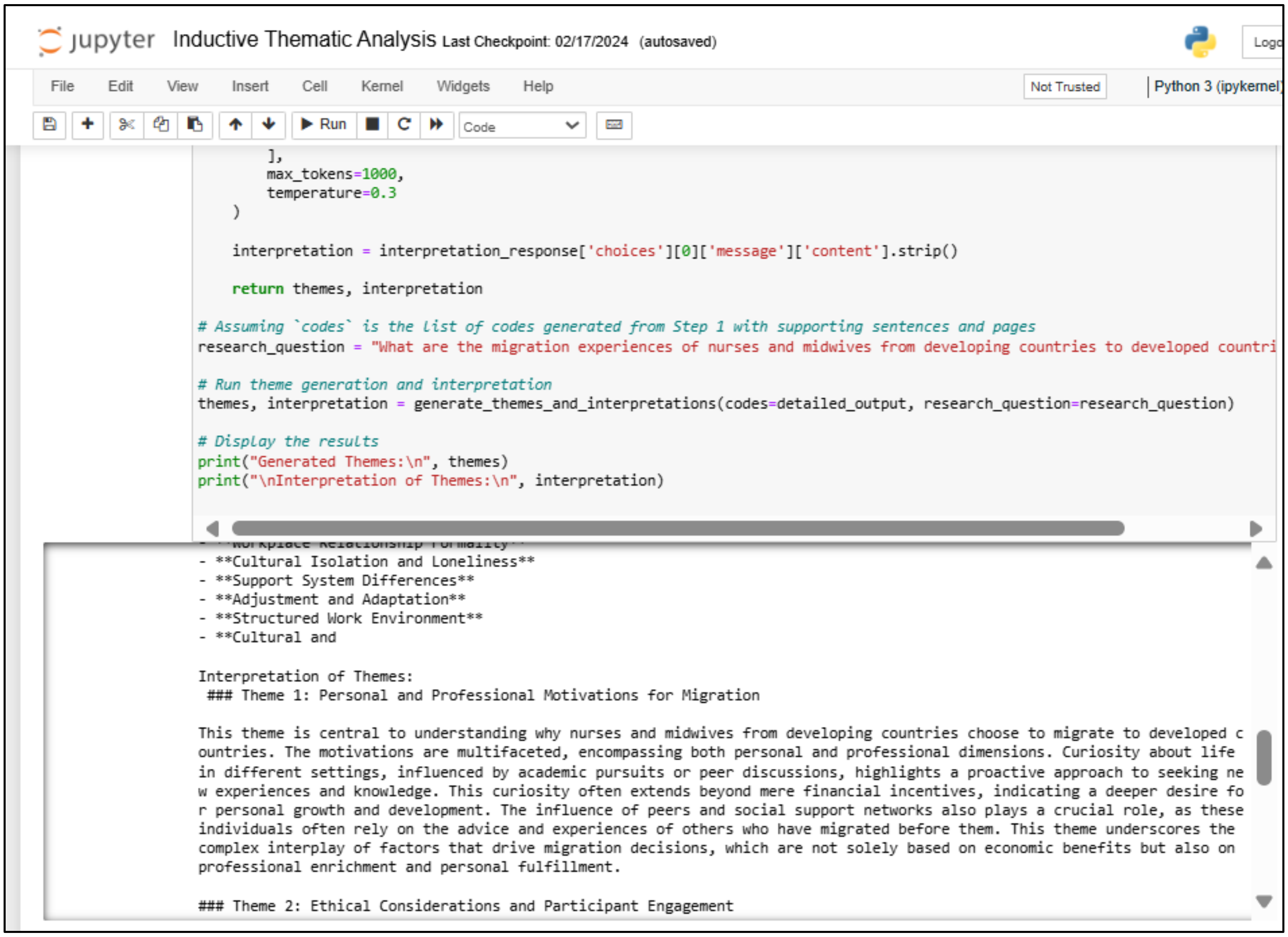

*Figure 5: Interpretation of the Data by GenAI*

**Human Coder**

Two qualitative researchers (Coder 1 and Coder2) conducted independent inductive coding, generating 69 and 102 codes, respectively. A third reviewer then integrated their results into a merged dataset of 106 human-derived codes (See Table 1). For instance, Coder 1 coded *Career Foundations* to reflect the respondent's professional background and early influences, while Coder 2 used the more detailed *Career Exploration* and *Career Formation* codes. Similarly, in terms of migration factors, Coder 1 used *Factors leading to Migration*, while Coder 2 used *Factors influencing Migration*, emphasizing a broader understanding of migration influences. Human 1 and Human 2 codes were merged sinking the difference and outliers to form 106 human generated codes compared with GenAI (See Appendix). These codes were merged based on similarities. Similar codes identified by both coders (67 codes) were counted as one, while unique (outlier) codes found in only one coder's analysis were counted separately and added to the total.

**Table 1**

*Frequency count of Codes by Human Coders*

| Coder | Number of Codes |
|---|---|
| Coder 1 | 69 |
| Coder 2 | 102 |
| Similar Codes | 67 |
| Merged Codes | 106 |

*Human Themes*

We further compared themes of the two coders, highlighting both consensus and divergent ideas regarding themes (See Table 2). The analysis of the themes process shows a moderate level of agreement between Coder 1 and Coder 2. Coder 1 identified 23 codes (65.22%) that overlapped with Coder 2's coding, while Coder 2 identified 26 (57.69%) that overlapped with Coder 1's coding. In total, both human coders agreed on 15 similar codes. Table 3 shows example themes of the two coders includes three main themes: *Professional / Career background*, *Rationale for midwife career*, and *Factors that influenced decision to migrate*. For instance, both reviewers agreed that the respondent's qualifications and prior experience allowed her to settle comfortably into the profession in the UK (Consensus Ideas for *Professional / Career background*). However, a key divergence arose in the coding of career-related experiences: Coder 1 used a broad "career history" code, while Coder 2 separated academic qualifications and work experience into distinct categories. Similarly, for the theme Factors that influenced decision to migrate, both reviewers agreed on factors like curiosity and peer influence but diverged on the emphasis of rejecting financial motivations, which Coder 2 included but not Coder 1.

**Table 2**

*Comparison of Code Similarity Between Coder 1 and Coder 2*

| Themes | Count | Percentage of Similarity |
|---|---|---|
| Coder 1 | 23 | 65.22% |
| Coder 2 | 26 | 57.69% |
| Similar codes between the two Coders | 15 | |

**Table 3** codes
*Example of Themes of the Two Human Coders*

| Theme | Main Ideas | Consensus Ideas | Divergent Ideas |
|---|---|---|---|
| Professional / Career background | This theme encompasses the respondent's previous work experience and academic qualifications. | Both reviewers agree that the respondent's training, qualification, and previous experiences as a midwife enabled her to settle comfortably into the profession in the UK. | The main divergence is in coding: Reviewer 1 uses "career history" as a broad code, whereas Reviewer 2 separates academic qualification and work experiences. |
| Rationale for midwife career | This theme traces the respondent's journey into the midwifery profession. | Both reviewers highlight that the respondent became a midwife by "chance" and the influence of family members on the applicant's journey into midwifery. | Reviewer 1 uses "motivation to pursue midwifery," while Reviewer 2 uses "career exploration" and "career formation," producing more nuanced codes such as "lack of career guidance" and "rediscovering passion." |
| Factors that influenced decision to migrate | This theme focuses on factors that influenced the respondent's decision to migrate to the UK. | The common ground is in the codes and illustrative quotes used to represent ideas such as "curiosity," "motivation from | Reviewer 2 includes "rejecting financial motivations" and the quote "money wasn't for me," which Reviewer 1 omits, emphasizing non- |

| | | |
|---|---|---|
| | friends or peer influence," and "discovery of truth." | financial reasons for migration. |

*Human Interpretations*

We further compared interpretation given by the two coders, highlighting both consensus and divergent ideas regarding themes. For instance, in the theme "Factors that Influence Decisions to Migrate," the main idea that ran through the interpretations of both coders was that "This theme emphasizes the multifaceted motivations behind migration decisions, highlighting both personal and professional factors that drive individuals to migrate." Both coders reached a consensus on this overarching theme, agreeing that people's qualifications, previous training, and experiences, particularly as midwives, play a key role in their ability to settle in a new environment. However, their interpretations diverged on specific aspects. While Coder 1 emphasized how the training and prior experiences enable the individual to settle comfortably into the new country, Coder 2 placed more emphasis on how these qualifications not only enable settling but also contribute to the individual's expertise, adaptability, and growth potential within the profession. This distinction between settling and professional growth showcased the divergence in their interpretations. Examples of the themes and interpretation include:

"Theme 1: Professional / Career background

*Main ideas: This theme encompasses the respondent's previous work experiences and academic qualifications*

*Consensus ideas and extended point: Both reviewers agree that the respondent's training, qualification and previous experiences as a midwife has enabled her settle comfortably into the profession in the UK. However, unlike reviewer 1, reviewer 2 emphasizes how the applicant's qualifications and prior experiences has contributed to her expertise, adaptability and potential for growth within the midwifery profession in the UK.*

*Divergent ideas and extended points: The main divergence between both reviewers is with respect to the coding. Reviewer one uses "career history" as a broad code to encompass the respondent's academic qualification and work experiences whereas reviewer two uses two separate codes for each of these.*

Theme 2: Rationale for midwife career

*Main ideas: This theme traces the respondent's journey into the midwifery profession.*

*Consensus ideas and extended point: Both reviewers highlight that the respondent became a midwife by "chance". This is reflected in the title of their codes and use of the illustrative*

*quote "I chance on midwifery, and I fell in love with it". Further, both reviewer's highlight the influence of family members on the applicant's journey into becoming a midwife.*

*Divergent ideas and extended points: The divergence between both reviewers occurs in terms of the themes and codes. Reviewer one uses the theme "motivation to pursue midwifery" to capture the respondent's journey into midwifery whiles reviewer two uses "career exploration" and "career formation" respectively to capture these ideas. As a result, reviewer two produces more nuanced codes under this theme than reviewer one. This is reflected in codes such as "lack of career guidance", "shifting career aspirations", "initial lack of interest" and "rediscovering passion".*

<u>Theme 3: Factors that influenced decision to migrate</u>

*Main ideas: This theme focuses on factors that influenced the respondent's decision to migrate to the UK.*

*Consensus ideas and extended point: The common ground between both reviewers is in terms of the codes and illustrative quotes used to represent these ideas. The code "curiosity" appears among both reviewers just as "motivation from friends or peer influence" as well as "discovery of truth or seeking truth".*

*Divergent ideas and extended points: The main difference between both reviewers is in terms of the code "rejecting financial motivations" and the illustrative quote "people will say money, but money wasn't for me" which is missing in reviewer one's codes. By so doing, reviewer two is able to emphasize that the respondent's decision to migrate was driven by other factors beyond money."*

## Comparing GenAI and Human

### *GenAI-Human Coding*

This analysis suggests that while there is a small difference in the number of codes generated, the overall percentage similarity indicates a high level of alignment between the outputs of the human coders and the GenAI, reflecting the GenAI's effectiveness in generating content similar to that of human coders (See Table 4). The human coders produced 53.17% (n=67) codes of the total 100% (n=126) codes, that is the combined codes between the average codes of human and GenAI codes. On the other hand, GenAI generated 46.83% (=59) codes of the total. This indicates that GenAI produced 11.94% (n=8) fewer codes than the human coders. The percentage similarity between the codes generated by the human coders and the GenAI is 88.06% (118) codes, which signifies a strong overlap. In other words, there is an 88.06% agreement between the codes produced by both the human coders and GenAI, highlighting that a large majority of the codes generated by both sources are similar in nature. For examples of the similar codes include, the themes *Accidental Career Discovery*, *Curiosity-Driven Migration, Peer Influence,*

*Delayed Access, and Merit-Based Opportunity* were all identified by Human Coders and GenAI. For each of these codes, rated with a 1, indicating present (See Table 5).

$$\text{Percentage Difference} = \frac{Human\ Codes - GenAI\ Codes}{Human\ Codes} \times 100$$

$$\text{Percentage Difference} = \frac{67 - 59}{67} \times 100 \approx 11.94\%$$

$$Percentage\ Similarity = 100 - 11.94, 88.06\%$$

**Table 4**

*Comparison of Code Counts Between Human Coders and GenAI Coders*

| Measure | Count | Percentage |
|---|---|---|
| Average Codes of Human (Similar Codes) | 67 | 53.17% |
| GenAI Codes | 59 | 46.83% |
| Percentage Difference | 8 | 11.94% less |
| Percentage Similarity | 118 | 88.06% |
| Total Average Codes of Human and GenAI Codes | 126 | 100% |

NB:

**Table 5**

*Example of Similar Codes Across Average Human Coders and GenAI*

| Code | Average Human | GenAI |
|---|---|---|
| Accidental Career Discovery | 1 | 1 |
| Curiosity-Driven Migration | 1 | 1 |
| Peer Influence | 1 | 1 |
| Delayed Access | 1 | 1 |
| Merit-Based Opportunity | 1 | 1 |

*NB: Nominal number 1 rate shows coded, and 0 rate shows not coded*

*GenAI-Human Themes*

The comparison between GenAI Themes and Average Themes (Similar of Human Coders) shows that GenAI Themes account for 21.05% (n=4) of the total themes when both are combined, while Average Themes (Similar of Human Coders) represent 78.95% (n=15) of the total themes. However, comparing the 15 human themes with GenAI list of emerging codes that can be considered as themes, gave us 100% count between the two (See Table 6). This indicates that a larger proportion of themes were identified by humans compared to the GenAI identified themes but equal proportion when compared with the GenAI list of emerging codes that can be considered as themes.

Example of when comparing the 15 average themes of the human coders with the 4 GenAI themes. The human coders' themes included *Professional / Career Background, Rationale for Midwife Career, and Factors that Influenced the Decision to Migrate*, while the 4 GenAI themes were broader, such as *Personal and Professional Motivations for Migration, Ethical Considerations, Career Trajectory, and Cultural and Social Adjustments.* These themes were similar in that both identified migration motivations, career development, and personal factors. However, they diverged in that the GenAI themes were more generalized and consolidated, while the human coders focused on specific factors like career background and motivation for migration. Next, when comparing the 15 average human themes with the 15 emerging codes generated by GenAI, which could be considered as themes. For example, GenAI's emerging codes, such as *Accidental Career Discovery, Curiosity-driven Migration, and Peer Influence on Migration Decision*, were like Human Coders' themes like *Curiosity-driven Migration and Peer Influence*. However, GenAI's emerging codes were more granular, and themes like *Accidental Career Discovery* and *Non-financial Motivations* were not explicitly identified in the human coders' themes. These themes show similarity in their recognition of key migration factors but diverge in how they categorize and group these factors.

**Table 6**

*Percentage Comparison of GenAI Themes and Average Themes of Human*

| Themes | Count | % Similarity of Qty |
|---|---|---|
| If, List of Emerging Codes as themes | 15 | 100% |
| GenAI Themes | 4 | 21.05% |
| Average Themes (Similar of Human Coders) | 15 | 78.95% |

*GenAI-Human Interpretations*

The comparison between the Human Coders' Interpretations and GenAI's Interpretation shows both alignment and divergence in the main ideas and consensus. In the theme "Factors that Influence Decisions to Migrate," both Human Coders and GenAI agree on the multifaceted nature of migration motivations, emphasizing both personal and professional factors. For instance, Human Coders reached a consensus that qualifications and prior training, particularly as midwives, are key to an individual's ability to settle into a new environment. This aligns with GenAI's interpretation, which also emphasizes the importance of personal and professional motivations for migration, noting that curiosity and peer influence play significant roles in driving migration decisions (See Table 7). While the Human Coders and GenAI share a common understanding of the motivations behind migration, the Human Coders' Interpretations are more specific, with a clear distinction between settling into the profession and professional growth. Coder 1 focused on how prior experiences help individuals settle, while Coder 2 highlighted how qualifications contribute to adaptability and professional growth. In contrast, GenAI combines these aspects into one broader theme, focusing on the proactive approach to migration driven by curiosity, social support, and personal development.

**Table 7**

*Comparison of Human Coders' Interpretations and GenAI Interpretation*

| Aspect | Human Coders' Interpretations | GenAI Interpretation |
|---|---|---|
| Main Idea | Focuses on the multifaceted motivations behind migration decisions, with specific emphasis on qualifications and training for settling and growth. | Highlights the personal and professional motivations for migration, with a focus on curiosity, peer influence, and personal development. |
| Consensus | Both coders agreed that qualifications and previous training, particularly as midwives, play a key role in migration decisions, enabling individuals to settle into a new environment. | GenAI agrees that migration decisions are influenced by both personal and professional motivations but presents a broader view combining multiple aspects into a single theme. |
| Divergence | Coder 1 emphasizes settling into the profession, while Coder 2 highlights professional growth and adaptability. | GenAI merges both settling and professional growth into a broader, generalized theme focused on proactive migration driven by curiosity and social support. |
| Overall Comparison | Human Coders provide detailed, specific themes that distinguish between settling and professional growth, while GenAI offers a more | GenAI's broader approach captures the complexity of migration but lacks the depth and nuance seen in the human |
| | generalized interpretation of the migration motivations. | coders' specific analysis of settling vs. growth. |

## Discussions

The GenAI inductive analysis, utilizing OpenAI with a stepwise prompting approach, demonstrated greater transparency and optimized thematic analysis compared to previous studies. For instance, Lee et al. (2024) found that while GenAI effectively generates themes, it may not fully capture the nuanced context of participant responses, suggesting that it serves better as a complementary tool rather than a replacement for human analysis in qualitative research. However, the findings of this study indicate that GenAI can accurately capture human responses with supporting sentences that enhance clarity and justification. Averagely, GenAI captured near 100% of codes and a 100% of themes which confirms and extends Prescott et al. (2024) and Qiao et al. (2024) studies indicating that GenAI has 71% and 80% accuracy in coding respectively. Moreover, this study reinforces the importance of transparency in GenAI-assisted analysis, addressing concerns about hallucinations, which De Paoli (2024) highlighted as a significant challenge in GenAI-driven qualitative research. The results further support Bijker et al. (2024), who found that GenAI performs particularly well with inductive coding schemes, confirming its effectiveness in systematically identifying themes in qualitative data.

Our analysis also revealed that GenAI's procedure closely aligns with the six-step thematic analysis framework of Clarke and Braun (2016) and, more specifically, with Naeem et al. (2023), even more so than the human coders in this study. Although the stepwise prompts did not explicitly outline Naeem et al.'s process, a clear connection emerges when mapping GenAI's workflow onto their framework. The codes generated by GenAI can be linked to Step 2 (Keywords) of Naeem et al. (2023), while its list of emerging codes aligns with Step 3 (Coding). Additionally, themes identified by GenAI correspond to Step 4 (Theme Identification), and its interpretations align with Step 5 (Conceptualization). Based on this alignment, we conclude that GenAI followed at least four of the six steps, demonstrating a structured and systematic thematic analysis approach. By contrast, if human coding were mapped onto Naeem et al.'s framework, highlighted words would correspond to Step 2 (Keywords), themes to Step 3 (Coding), and interpretations to Step 4 (Themes). This finding challenges Bijker et al. (2024), who suggested that GenAI might struggle to follow existing data analysis frameworks. Even though the process here was inductive rather than deductive, our results suggest that GenAI employs an algorithmic approach that can closely adhere to established thematic analysis procedures, reinforcing its potential for structured qualitative research.

The stepwise prompts used in script generation produced a highly structured output, effectively categorizing data into codes, a list of emerging codes, themes, and interpretations with notable accuracy. Both GenAI and human coders utilized the same codes and themes, but GenAI demonstrated a more standardized and broad interpretative approach, whereas human coders provided more specific and contextually aligned interpretations. This suggests that human analysis leaned more toward theme-based procedures, aligning with Step 4 (Themes) in Naeem et al. (2023), while GenAI extended Step 5 (Conceptualization) by offering broader generalizations. Consequently, GenAI exhibited comparable proficiency in keywording and coding but demonstrated a higher-level conceptualization approach than human coders, whose interpretations remained more tightly linked to codes and themes. This supports Bijker et al. (2024), who emphasized the importance of prompt engineering in guiding ChatGPT for qualitative data analysis, reinforcing its ability to follow structured frameworks effectively.

Based on our experience in this study, we consider GenAI's API call as a valuable tool for assisting researchers in conducting inductive thematic analysis with a level of accuracy comparable to human expertise. GenAI effectively executes keywording, coding, theming, and interpretation, closely aligning with structured qualitative methodologies. This finding contrasts sharply with Morgan (2023), who found that while GenAI successfully identified basic thematic material, it struggled with capturing subtle, interpretive themes. However, our study suggests that the accuracy of this approach surpasses the baseline capabilities of general-purpose GPTs, reducing the need for extensive revisions in some steps of the inductive thematic analysis process. We also acknowledge that using API-based approaches may be more technically demanding than working with base GPTs that function through simpler text-based interactions.

However, we believe that integrating this method into qualitative data analysis could become a more accessible learning approach for researchers, like how they adapt to Computer-Assisted Qualitative Data Analysis Software (CAQDAS). Dahal (2024) predicted a future where GenAI would play a significant role in qualitative research as a co-author, research assistant, and conversational tool, raising both opportunities and ethical concerns. However, our findings suggest that this future is already here, as GenAI in our study successfully performed all research stages, from coding and theming to interpretation, demonstrating its capability to enhance transparency, accuracy, and efficiency in inductive thematic analysis.

The findings also suggests that GenAI can be useful for coding and analysing qualitative data as long as the transcribed data is provided (Akbar, Khan & Liang, 2023; De Paoli, 2023; Gao et.al., 2023; Rahman & Watanobe, 2023; Sallam, 2023; Xiao et.al, 2023). So far as this condition is met, Akbar et. al., (2023) observes that ChatGPT can derive codes, identity key ideas, patterns and categories from the transcripts. These assertions have been confirmed by De Paoli (2023) who conducted an experiment to test the viability of using GAI tools for inductive thematic analysis. Irrespective of the positive results realized, De Paoli (2023) recommends the need to consider the methodological implications of using same for qualitative analysis whiles Akbar et.al., (2023) also cautions against the possible validity or credibility threats therein (Acheampong & Nyaaba, 2024)

**Conclusion and Future Direction**

This study highlights the transparency and accuracy of GenAI's inductive thematic analysis, particularly using GPT-4 Turbo API integrated within a Stepwise Prompt-based Python Script. This approach ensured a traceable and systematic coding process, generating codes with supporting statements and page references, which enhanced validation and reproducibility. The results indicate that GenAI performs inductive coding in a manner closely resembling human coders, effectively categorizing themes at a level like the average human coder. However, in interpretation, GenAI extends beyond human coders by situating themes within a broader conceptual context, providing a more generalized and abstract perspective.

When we further align both GenAI and Human coding processes with Naeem et al.'s (2023) six-step systematic framework, GenAI appeared to have followed four out of the six steps, while human coders followed three. Both processes are aligned with the systematic framework in keywording, coding, and theming. However, the key difference is that the GPT-4 Turbo API-based approach also incorporated the interpretation step, which involves analyzing the meaning and significance of the themes identified, whereas it was not explicitly part of the human-based method. These findings position GenAI, via OpenAI's API, as a promising tool for optimizing inductive thematic analysis, potentially reducing or even eliminating the need for human review in certain steps. We recommend that future studies explore the use of systematic frameworks for inductive coding as structured prompts or models for GenAI-driven thematic analysis. We also acknowledge that using API calls for GenAI may present challenges for researchers with limited coding experience, making the chatbot version with plain language prompts a more accessible alternative. Therefore, we recommend adopting a stepwise prompting approach within chatbot interfaces and iterating through responses to assess its effectiveness in achieving similar outcomes as the API-based approach.

**Appendices**

Appendix 1:

*GenAI and OpenAI API Analysis of the Interview Data*

| Theme | Codes | Description | Interpretation |
|---|---|---|---|
| Theme 1: Personal and Professional Motivations for Migration | Academic Background of Researcher, Migration Focus, Curiosity-driven Migration, Non-financial Motivation, Peer Influence on Migration Decision, Motivation Through Social Support, Motivation Beyond Financial Gain | This theme explores personal and professional reasons motivating individuals to migrate, including curiosity about conditions abroad, academic pursuits, peer influences, and intrinsic motivations beyond financial gains. | This theme is central to understanding why nurses and midwives from developing countries choose to migrate. It captures multifaceted motivations, with curiosity, academic pursuits, and peer influence playing crucial roles in migration decisions. The theme highlights a proactive approach toward personal growth and knowledge, indicating that motivations often extend beyond financial incentives. |
| Theme 2: Ethical Considerations and Participant Engagement | Confidentiality Assurance, Permission for Participation | This theme addresses ethical considerations, emphasizing confidentiality and informed consent in engaging | This theme reflects the ethical foundation necessary for research involving migrant nurses and midwives, with |
| | | participants for studies. | confidentiality and informed consent crucial to trust and credibility. The theme underscores broader ethical considerations relevant to migration experiences, such as respect and fair treatment. Ethical practices in research mirror the professional standards these individuals expect, affecting their willingness to engage in studies and their sense of value within professional settings. |
| Theme 3: Career Trajectory and Professional Development | Accidental Career Discovery, Influence of Childhood Experience, Shift in Career Aspiration, Family Influence on Career Choice, Decision to Pursue Midwifery, Successful Admission and Career Fulfillment, Long Wait Times for Professional Advancement, Financial Challenges | This theme captures the evolution of career aspirations, personal influences, and opportunities or challenges related to professional growth, both domestically and abroad. | Career development is a significant theme among migrating nurses and midwives, with many seeing migration as an opportunity to advance beyond what's possible in their home countries. This includes access to training, technologies, and |
| | in Professional Development, Merit-based Opportunity in the UK, Direct Support for Professional Growth in the UK, Readiness and Relevance Over Seniority, Professional Development Opportunities, Career Progression Impact, Desire for Professional Development, Underutilization of Skilled Workers | | professional networks. However, this theme also encompasses challenges such as long wait times for advancement and financial barriers. Merit-based systems in places like the UK offer transparency and career growth, illustrating how migration provides substantial opportunities but also requires resilience and adaptation. |
| Theme 4: Cultural and Social Adjustments | Initial Cultural Shock, Unexpected Weather Conditions, Quiet and Orderly Environment, Healthcare System Accessibility and Procedures, Delayed Medical Services and Payments, Equal Treatment in Healthcare, Workplace Relationship Formality, Cultural Isolation and Loneliness, Support System Differences, Adjustment and | This theme captures cultural and social adjustments that migrant nurses and midwives experience in adapting to new environments. | Cultural and social adjustments are crucial in the migration journey for nurses and midwives, with initial cultural shocks, new social norms, and workplace adjustments requiring significant adaptation. This theme underscores the need for support systems to ease transitions and |

| | | |
|---|---|---|
| Adaptation, Structured Work Environment | | highlights resilience and adaptability among migrant professionals. Adjusting to unfamiliar weather, healthcare procedures, and workplace formality often impacts both personal well-being and professional satisfaction, making these themes essential for understanding the broader migration experience. |

**Appendix 2**

*Human-generated themes vs. AI-generated themes*

| Human Reviewer 1 (n=23) | Human Review 2 (n=26) | Key Consensus Themes | Ai Themes (N=4) | Divergent themes |
|---|---|---|---|---|
| Professional background | Career Foundations | | Career Trajectory and Professional Development | |
| Motivation to pursue Midwifery | Career Exploration | | | |
| | Career Formation | | | |
| Factors that influenced migration | Factors leading to Migration | | Personal and Professional Motivations for Migration | |
| Professional opportunities in Ghana, compared with UK | Comparative Career Opportunities: Ghana vs the UK | | | |
| Source of information about migration | Sources of Information | | Cultural and Social Adjustments | |
| Challenges in accessing information | Navigating the Information Maze | | | |
| Certification and additional training during migration | Pathways to UK Midwifery QUalification | | | |
| Documentation and visa process | Documentation and Visa Process Support | | | |
| Support during migration | Migration Support | | | |

| | |
|---|---|
| Family and friends reaction towards migration | Family Emotional and Financial Dynamics of Migration |
| Support network for Nurses & Midwives | Support Systems for Nurse and Midwife Migration |
| Initial impressions about new country | First Impressions: Encountering a New Environment |
| Adaptation to new healthcare system | Adapting to the UK healthcare system |
| Significant challenges during adjustment period | Challenges during Adjustment Period |
| Comparison of working conditions in home country and new country | Contrasting Workplace Cultures: Ghana and the UK |
| Professional Development Opportunities encountered | Reality of Professional Development Opportunities |
| Job satisfaction | Job Satisfaction in the UK |
| Life challenges during migration | Financial Realities and Job Satisfaction in the UK |
| Impact of migration on personal life and wellbeing | Impact of Migration: Adapting and Thriving |
| Family & Social relationship | Shifting Relationships: |

| | | |
|---|---|---|
| | Family Bonds and Evolving Friendships | |
| Future plans | Longing for Home: Future Aspirations and Sense of Belonging | |
| Advice to nurses and midwives | Migration: A Journey of Personal Growth and Informed Choices | |
| | Optimising Ghana's Healthcare Workforce | |
| | Overcoming Obstacles | |
| | Independent Pathways to Migration | |
| Suggestions for improving migration in Ghana | | |
| | | Ethical Considerations and Participant Engagement |

| | | | | | | | | | | |
|---|---|---|---|---|---|---|---|---|---|---|
| Merged codes (Either Human-merged or GenAI) N=115<br><br>120 | A. Merged (All) codes (Human 1 Ư Human 2) N=108 :106 | B. Human Coder 1 Only (All codes)<br><br>N=69 | C. Human Coder 2 Only (All codes)<br><br>N=106 | D. Similar Human Codes (Coder 1 ∩ Coder) N= | E. Gen-AI Only Codes N=59 | F. Similar Codes of Human Codes (Human Coder 1 ∩ Human Coder) ∩ GenAI Codes<br><br>D ∩ E<br><br>N= | Outlier codes (Different Code between GenAI and Similar Codes of Human Codes (Human Coder 1 ∩ Human Coder)<br><br>N=<br><br>(*Indicate if GenAI or Human) | Similar codes between GenAI and merged codes by human<br><br>N=42 | Outlier codes (Different Code between GenAI and merged codes by human)<br><br>N=80 | Merged codes (Either Human-merged or GenAI) N=115 |
| 1. 1.Work experiences | 1. Work experiences | 1 | 1 | 1 | | | | | Work experiences | 1.Work experiences |
| 2. 2.Academic qualification | 2. Academic qualification | 1 | 1 | 1 | | | | 71. | Academic qualification | 2.Academic qualification |
| 3. 3.Accidental Career Discovery | 3. Career Midwife by chance | 1 | 1 | 1 | 5.Accidental Career Discovery | | | Accidental Career Discovery | | 3.Accidental Career Discovery |
| 4. 4.Family Influence on Career Choice | 4. Family influence on career | 1 | 1 | 1 | 8.Family Influence on Career Choice | | | Family Influence on Career Choice | | 4.Family Influence on Career Choice |
| 5. 5. Lack of career guidance | 5. Lack of career guidance | 1 | 1 | 1 | | | | | Lack of career guidance | 5. Lack of career guidance |
| 6. 6.Shift in Career Aspiration | 6. Shifting career aspirations | 0 | 1 | | 7.Shift in Career Aspiration | | | Shift in Career Aspiration | | 6.Shift in Career Aspiration |
| 7. 7.Initial lack of interest | 7. Initial lack of interest | | 1 | | | | | | Initial lack of interest | 7.Initial lack of interest |
| 8. 8.Rediscovering passion | 8. Rediscovering passion | 0 | 1 | | | | | | Rediscovering passion | 8.Rediscovering passion |
| . Influence of Childhood Experience (Ai) | 0 | 0 | 0 | | 6. Influence of Childhood Experience | | | | 6. Influence of Childhood Experience (Ai) | . Influence of Childhood Experience (Ai) |
| 9. 9. Decision to Pursue Midwifery (Ai) | | | | | 9. Decision to Pursue Midwifery | | | | 9. Decision to Pursue Midwifery (Ai) | 9. Decision to Pursue Midwifery (Ai) |
| 10. 9.Successful Admission and Career Fulfilmen | 0 | 0 | 0 | | 10.Successful Admission and Career Fulfilment | | | | 10.Successful Admission and Career Fulfilmen | 9.Successful Admission and Career Fulfilmen |
| 11. 10.Curiosity-driven migration | 9. Curiosity | 1 | 1 | 1 | 11.Curiosity-driven Migration | | | Curiosity-driven migration | | 10.Curiosity-driven migration |
| 12. 11.Pursuit of truth | 10. Pursuit of truth | 1 | 1 | 1 | | | | | Pursuit of truth | 11.Pursuit of truth |
| 13. 12. Non-financial Motivation | 11. Rejecting financial motivation | | 1 | | 12.Non-financial Motivation | | | .Non-financial Motivation | | 12. Non-financial Motivation |
| 14. 13. Peer influence on migration decision | 12. Peer influence | 1 | 1 | 1 | 13.Peer Influence on Migration Decision | | | Peer influence on migration decision | | 13. Peer influence on migration decision |
| 15. 14. Perceived Barriers to Migration | 0 | 0 | 0 | | 14. Perceived Barriers to Migration | | | | 14. Perceived Barriers to Migration (Ai) | 14. Perceived Barriers to Migration |
| 16. 15. Delayed access to professional development and | 13. Delayed access to professional opportunities and | 1 | 1 | 1 | 16. Long Wait Times for Professional Advancement | | | Delayed access to professional development and | | 15. Delayed access to professional development and |
| opportunities in Ghana | career growth in Ghana | | | | | | | opportunities in Ghana | | opportunities in Ghana |
| 17. 16. Lengthy progression process in Ghana | 14. Lengthy progression process in Ghana | 0 | 1 | | | | | | Lengthy progression process in Ghana | 16. Lengthy progression process in Ghana |
| 18. 17.Faster career progression in UK | 15. Accelerated progression process in the UK | 1 | 1 | 1 | | | | | Faster career progression in UK | 17.Faster career progression in UK |
| 19. 18 Financial challenges in Ghana | 16. Financial challenges in Ghana | 1 | 1 | 1 | 17. Financial Challenges in Professional Development | | | Financial challenges in Ghana | | 18 Financial challenges in Ghana |
| 20. 19.Fairness in UK's system | 17. Fairness in UK's system | 1 | 1 | 1 | | | | | Fairness in UK's system | 19.Fairness in UK's system |
| 21. 20.Merit-based Opportunity in the UK | 18. Merit based system in the UK | 1 | 1 | 1 | 18. Merit-based Opportunity in the UK | | | Merit-based Opportunity in the UK | | 20.Merit-based Opportunity in the UK |
| 22. 21.Financial support for professional developme | 19. Financial support available in the UK | 1 | 1 | 1 | 19. Direct Support for Professional Growth in the UK | | | Financial support for professional development in the UK | | 21.Financial support for professional development in the UK |

| | | | | | | | | | |
|---|---|---|---|---|---|---|---|---|---|
| nt in the UK | | | | | | | | | |
| 23. 22.Readiness and Relevance Over Seniority | 0 | 0 | 0 | | 20.Readiness and Relevance Over Seniority | | | | Readiness and Relevance Over Seniority (Ai) | 22.Readiness and Relevance Over Seniority |
| 24. 23.Independent information gathering | 20. Independent information gathering | 1 | 1 | 1 | | | | | Independent information gathering | 23.Independent information gathering |
| 25. 24.Diverse information sources | 21. Diverse information sources | 1 | 1 | | | | | | Diverse information sources | 24.Diverse information sources |
| 26. 25. Role of Internet in Migration Information Gathering | 22. Internet cafe | 1 | 1 | 1 | 25. Role of Internet in Migration Information Gathering | | | Role of Internet in Migration Information Gathering | | 25. Role of Internet in Migration Information Gathering |
| 27. 26.Strangers and colleague | 23. Strangers and colleagues | 1 | 1 | 1 | | | | | Strangers and colleagues | 26.Strangers and colleague |
| 27. Challenges in accessing Reliable resources | 24. Difficulty in getting authentic information | 1 | 1 | 1 | 24.Challenges in Accessing Reliable Resources | | | Challenges in accessing Reliable resources for | 23.Verification of Information (Ai) | 27. Challenges in accessing Reliable resources for immigration |
| for immigration<br><br>28. 3.Verification of Information (Ai) | | | | | 23.Verification of Information | | | immigration | | 3.Verification of Information (Ai) |
| 29. 28.Difficulty finding learning resources | 25. Difficulty finding learning resources | 1 | 1 | 1 | | | | | Difficulty finding learning resources | 28.Difficulty finding learning resources |
| 30. 29.Lack of transparency and openness | 26. Lack of transparency and openness | 1 | 1 | 1 | | | | | Lack of transparency and openness | 29.Lack of transparency and openness |
| 31. 30.Information overload and reliability | 27. Information overload and reliability | 1 | 1 | 1 | | | | | Information overload and reliability | 30.Information overload and reliability |
| 32. UK Certification and Testing Requirement | 28. UK qualification requirement | 1 | 1 | 1 | 26. Certification and Testing Requirements | | | UK Certification and Testing Requirement | | UK Certification and Testing Requirement |
| 33. English language proficiency test and Computer- | 29. English language proficiency test and | 1 | 1 | 1 | | | | | English language proficiency test and Computer- | English language proficiency test and Computer- |
| based test on area of specialisation | Computer-based test on area of specialisatio | | | | | | | | based test on area of specialisation (H) | based test on area of specialisation |
| 34. 32.Securing sponsorship | 30. Securing sponsorship | 0 | 1 | 1 | | | | | Securing sponsorship (H) | 32.Securing sponsorship |
| 35. 33.Practical Examinations in Destination Country | 31. Registration and Licensing | 1 | 1 | 1 | Practical Examinations in Destination Country | | | Practical Examinations in Destination Country | | 33.Practical Examinations in Destination Country |
| 36. 34.Support from Partner and Employer<br>37. Supported by hospital | 32. Support from spouse | 1 | 1 | 1 | 30. Support from Partner and Employer | | | Support from Partner and Employer | | 34.Support from Partner and Employer<br>Supported by hospital |
| 0 | 33. Supported by hospital | 1 | 1 | 1 | | | | | Supported by hospital | |
| 38. 35. Self-Reliance in Documentation and Visa Process | 34. Self-reliance and resourcefulness | 1 | 1 | 1 | 28. Self-Reliance in Documentation and Visa Process | | | Self-Reliance in Documentation and Visa Process | | 35. Self-Reliance in Documentation and Visa Process |
| 39. 36.Avoidance of Financial Burden from Agencies | 0 | 0 | 0 | | 29. Avoidance of Financial Burden from Agencies | | | | Avoidance of Financial Burden from Agencies (Ai) | 36.Avoidance of Financial Burden from Agencies |
| 40. 37.Comprehensive financial support from hospital | 35. Comprehensive financial support from hospital | 1 | 1 | 1 | 30. Support from Partner and Employer | | | Comprehensive financial support from hospital | | 37.Comprehensive financial support from hospital |
| 41. 38.Financial reimbursement of expenses | 36. Reimbursement of expenses | 1 | 1 | 1 | 31. Financial Burden and Reimbursement | | | Financial reimbursement of expenses | | 38.Financial reimbursement of expenses |
| 42. 39.Family Emotional Response to Migration | 37. Mixed emotions | 0 | 1 | | 32. Family Emotional Response to Migration | | | Family Emotional Response to Migration | | 39.Family Emotional Response to Migration |
| 43. 40. Family expectations and financial remittance | 38. Family expectations and financial remittance | 0 | 1 | | | | | | Family expectations and financial remittance | 40. Family expectations and financial remittance |
| 44. 41.Family Emotional Response to Migration | 39. Joyful news on transition | 1 | 1 | 1 | 32. Family Emotional Response to Migration | | | Family Emotional Response to Migration | | 41.Family Emotional Response to Migration |
| 45. 42. Challenges of Distance in Migration | 40. Sadness due to distance | 1 | 1 | | 33. Challenges of Distance in Migration | | | Challenges of Distance in Migration | | 42.Challenges of Distance in Migration |
| 46. 43.Acceptance and Positivity | 41. Acceptance and Positivity | 0 | 1 | | | | | | Acceptance and Positivity | 43.Acceptance and Positivity |
| 47. 44.Role of Online Support Networks | 42. Online support systems | 1 | 1 | 1 | 34. Role of Online Support Networks | | | Role of Online Support Networks | | 44.Role of Online Support Networks |
| 48. 45.Shared learning material | 43. Shared learning materials | 1 | 1 | 1 | | | | | Shared learning material | 45.Shared learning material |
| 49. 46.Advertised job opportunities | 44. Advertised job opportunities | 1 | 1 | 1 | | | | | Advertised job opportunities | 46.Advertised job opportunities |
| 50. 47. Quiet and Orderly Environment | 45. Beautiful environment | 1 | 1 | 1 | 37. Quiet and Orderly Environment | | | Quiet and Orderly Environment | | 47. Quiet and Orderly Environment |
| | 46. Quiet atmosphere | 1 | 1 | 1 | | | | | | |
| 51. 48.Initial Cultural Shock | 47. Cultural shock | 1 | 1 | 1 | 35. Initial Cultural Shock | | | Initial Cultural Shock | | 48.Initial Cultural Shock |
| 52. 49.Unexpected Weather Conditions | 48. Unexpected weather | 0 | 1 | | 36. Unexpected Weather Conditions | | | Unexpected Weather Conditions | | 49.Unexpected Weather Conditions |
| 53. 50.Good accessibility of healthcare in Ghana | 49. Good accessibility of healthcare in Ghana | 0 | 1 | | | | | | Good accessibility of healthcare in Ghana | 50.Good accessibility of healthcare in Ghana |
| 54. 51.Healthcare System Accessibility and Procedures in UK | 50. More restricted health system in the UK requiring appointments and referral | 1 | 1 | 1 | 38. Healthcare System Accessibility and Procedures | | | Healthcare System Accessibility and Procedures in UK | | 51.Healthcare System Accessibility and Procedures in UK |
| 55. Delayed Medical Services and Payments | | 0 | | | 39. Delayed Medical Services and Payments | | | | Delayed Medical Services and Payments | Delayed Medical Services and Payments |
| 56. 38. Healthcare System Accessibility and Procedures | 51. Limited weekend and holiday service in the UK | 1 | 1 | | 38. Healthcare System Accessibility and Procedures | | | 38. Healthcare System Accessibility and Procedures | | 38. Healthcare System Accessibility and Procedures |
| 57. 52.Cost of Prescriptions in the UK | 52. Cost of Prescriptions in the UK | 0 | 1 | 1 | | | | | Cost of Prescriptions in the UK | 52.Cost of Prescriptions in the UK |
| 58. 53.Prioritising health workers in Ghana | 53. Prioritising health workers in Ghana | 0 | 1 | | | | | | Prioritising health workers in Ghana | 53.Prioritising health workers in Ghana |
| 59. 54.Confidentiality and data protection in UK | 54. Confidentiality and data protection | 0 | 1 | | | | | | Confidentiality and data protection in UK | 54.Confidentiality and data protection in UK |
| 60. 55. Equal Treatment in Healthcare | 55. Everyone is treated equally in the UK | 1 | 1 | 1 | 40. Equal Treatment in Healthcare | | | Equal Treatment in Healthcare | | 55. Equal Treatment in Healthcare |
| 61. 56.Have general practitioners that provide basic care | 56. Have general practitioners that provide basic care | 1 | 1 | 1 | | | | | Have general practitioners that provide basic care | 56.Have general practitioners that provide basic care |
| 62. 57. Cultural shock | 57. Cultural shock | 0 | 1 | | | | | | Cultural shock | 57. Cultural shock |
| 63. 58.Cultural Isolation and Loneliness | 58. Isolation and loneliness | 0 | 1 | | 42. Cultural Isolation and Loneliness | | | Cultural Isolation and Loneliness | | 58.Cultural Isolation and Loneliness |
| 64. 59.Workplace Relationship Formality | | 0 | 0 | | 41. Workplace Relationship Formality | | | | Workplace Relationship Formality | 59.Workplace Relationship Formality |
| 65. 60.Adjustment and Adaptation | | 0 | 0 | | 44. Adjustment and Adaptation | | | | Adjustment and Adaptation | 60. Adjustment and Adaptation |
| 66. 61.Boredom before husband joined he | 59. Boredom before husband joined he | 1 | 1 | 1 | | | | | Boredom before husband joined he | 61.Boredom before husband joined he |
| 67. 62. Support System Difference | 60. Lack of communal support in the UK | 0 | 1 | | 43. Support System Difference | | | Support System Difference | | 62. Support System Difference |
| 68. 63.Poor working conditions in Ghana | 61. Poor working conditions in Ghana | 1 | 1 | 1 | | | | | Poor working conditions in Ghana | 63.Poor working conditions in Ghana |
| 69. 64.Inconsistent working hours in Ghana | 62. Inconsistent working hours in Ghana | 1 | 1 | 1 | | | | | Inconsistent working hours in Ghana | 64.Inconsistent working hours in Ghana |
| 70. 65.Structured Work Environment in UK | 63. Strict working hours in the UK | 1 | 1 | 1 | 45. Structured Work Environment | | | Structured Work Environment in UK | | 65.Structured Work Environment in UK |
| 71. 66.Paid and unpaid breaks in UK | 64. Paid and unpaid breaks in UK | 1 | 1 | 1 | | | | | Paid and unpaid breaks in UK | 66.Paid and unpaid breaks in UK |
| 72. 67.Collected Soap and Dettol from clients in Ghana | 65. Collected Soap and Dettol from clients in Ghana | 1 | 1 | 1 | | | | | Collected Soap and Dettol from clients in Ghana | 67.Collected Soap and Dettol from clients in Ghana |
| 73. 68.Restriction on taking gifts from patients in UK | 66. Restriction on taking gifts from patients in UK | 1 | 1 | 1 | | | | | Restriction on taking gifts from patients in UK | 68.Restriction on taking gifts from patients in UK |
| 74. 69.Lack of Christmas bonus in the UK | 67. Lack of Christmas bonus in the UK | 0 | 1 | | | | | | Lack of Christmas bonus in the UK | 69.Lack of Christmas bonus in the UK |
| 75. 70. Professional Development Opportunities | 68. Access to developmental opportunities in UK | 1 | 1 | 1 | 4 6. Professional Development Opportunities | | | Professional Development Opportunities | | 70. Professional Development Opportunities |
| 76. 71. Career Progression Impact | 69. Accelerated career advancement in the UK | 0 | 1 | | 47. Career Progression Impact | | | . Career Progression Impact | | 71. Career Progression Impact |
| 77. 72.Complex Job Satisfaction | 70. Lack of job satisfaction in UK | 1 | 1 | 1 | 48. Complex Job Satisfaction | | | Complex Job Satisfaction | | 72.Complex Job Satisfaction |
| 78. 73.Good salary but high cost of living | 71. Good salary but high cost of living | 1 | 1 | 1 | 50. Financial Misestimation | | | Good salary but high cost of living | | 73.Good salary but high cost of living |
| 79. 74.Comparative | 72. Relative improvement | 0 | 1 | | 49. Comparative | | | Comparative | | 74.Comparative Improvemen |
| Improvemen | | | | | Improvement | | | Improvement | | |
| 80. 75.Unexpected expenses | 73. Unexpected expenses | 0 | 1 | | | | | | Unexpected expenses | 75.Unexpected expenses |
| 81. 76.Financial challenges | 74. Financial challenges | 1 | 1 | 1 | | | | | Financial challenges | 76.Financial challenges |
| 82. 77.Manageability with budgeting | 75. Manageability with budgeting | 1 | 1 | 1 | | | | | Manageability with budgeting | 77.Manageability with budgeting |
| 83. 78.Increased Time Consciousness | 76. Increased time consciousness | 1 | 1 | 1 | 51. Increased Time Consciousness | | | Increased Time Consciousness | | 78.Increased Time Consciousness |
| 84. 79. Cultural and Dietary Adjustments | | 0 | 0 | | 52. Cultural and Dietary Adjustments | | | | Cultural and Dietary Adjustments (Ai) | 79. Cultural and Dietary Adjustments |
| 85. 80.Paid according to the number of hours worked in UK | 77. Paid according to the number of hours worked in UK | 1 | 1 | 1 | | | | | Paid according to the number of hours worked in UK | 80.Paid according to the number of hours worked in UK |
| 86. 81.Casual approach to | 78. Casual approach to | 0 | 1 | | | | | | Casual approach to | 81.Casual approach to |
| punctuality in Ghana | punctuality in Ghana | | | | | | | | punctuality in Ghana | punctuality in Ghana |
| 87. 82.Strict punctuality in the UK | 79. Strict punctuality in the UK | 1 | 1 | 1 | | | | | Strict punctuality in the UK | 82.Strict punctuality in the UK |
| 88. 83.Lifestyle changes | 80. Lifestyle changes | 1 | 1 | 1 | | | | | Lifestyle changes | 83.Lifestyle changes |
| 89. 84.Difficult work-life balance | 81. Difficult work-life balance | 0 | 1 | | | | | | Difficult work-life balance | 84.Difficult work-life balance |
| 90. 85.Increased Awareness of Mental Health | 82. Critical mental health awareness in the UK | 1 | 1 | 1 | 54. Increased Awareness of Mental Health | | | Increased Awareness of Mental Health | | 85.Increased Awareness of Mental Health |
| 91. 86.Physical and Mental Health Challenges | 83. Physical demands of the job in the UK | 1 | 1 | 1 | 53. Physical and Mental Health Challenges | | | Physical and Mental Health Challenges | | 86.Physical and Mental Health Challenges |
| 92. 87.Prioritising self-care | 84. Prioritising self-care | 1 | 1 | 1 | 56. Self-care Strategies | | | Prioritising self-care | | 87.Prioritising self-care |
| 93. 88.Work Intensity Differences | | 0 | 0 | | 5 5. Work Intensity Differences | | | | Work Intensity Differences (Ai) | 88.Work Intensity Differences |
| 94. 89.Enhanced Family Communication | 85. Strengthened family | 1 | 1 | 1 | 58. Enhanced Family | | | Enhanced Family Communication | | 89.Enhanced Family Communication |
| | connection | | | | Communication | | | | | |
| 95. 90.Isolation from Family Support | | 0 | 0 | | 57. Isolation from Family Support | | | | Isolation from Family Support (ai) | 90.Isolation from Family Support |
| 96. 91.Re-evaluation of friendships | 86. Re-evaluation of friendships | 0 | 1 | | | | | | Re-evaluation of friendships | 91.Re-evaluation of friendships |
| 97. 92.Financial request and expectation | 87. Financial request and expectation | 1 | 1 | 1 | | | | | Financial request and expectation | 92.Financial request and expectation |
| 98. 93.Changing Friendships and Exploitation | 88. Genuine vs Opportunistic friends | 0 | 1 | | 59. Changing Friendships and Exploitation | | | Changing Friendships and Exploitation | | 93. Changing Friendships and Exploitation |
| 99. 94.Lack of empathy and understanding | 89. Lack of empathy and understanding | 0 | 1 | | | | | | Lack of empathy and understanding | 94.Lack of empathy and understanding |
| 100.95.Thoughts on returning to Ghana | 90. Thoughts on returning to Ghana | 1 | 1 | 1 | | | | | Thoughts on returning to Ghana | 95.Thoughts on returning to Ghana |
| 101.96.Factors influencing return | 91. Factors influencing return | 1 | 1 | | | | | | Factors influencing return | 96.Factors influencing return |
| 102.97.Sense of belonging in Ghana | 92. Sense of belonging in Ghana | 0 | 1 | | | | | | Sense of belonging in Ghana | 97.Sense of belonging in Ghana |
| 103.98.Don't migrate solely for mone | 93. Don't migrate solely for money | 1 | 1 | 1 | | | | | Don't migrate solely for money | 98.Don't migrate solely for mone |
| 104.99.Follow your own pat | 94. Follow your own path | 1 | 1 | 1 | | | | | Follow your own pat | 99.Follow your own pat |
| 105.100.Migration should be a personal choice | 95. Migration should be a personal choice | 1 | 1 | 1 | | | | | Migration should be a personal choice | 100.Migration should be a personal choice |
| 106.101.Importance of personal growth | 96. Importance of personal growth | 1 | 1 | 1 | | | | | Importance of personal growth | 101.Importance of personal growth |
| 107.102.Benefits of travel | 97. Benefits of travel | 0 | 1 | | | | | | Benefits of travel | 102.Benefits of travel |
| 108.103.Travel as an alternative to migration | 98. Travel as an alternative to migration | 0 | 1 | | | | | | Travel as an alternative to migration | 103.Travel as an alternative to migration |
| 109.104.Flexible | 99. Flexible verification fees | 0 | 1 | | | | | | Flexible verification fees | 104.Flexible verification fees |
| verification fees | | | | | | | | | | |
| 110.105. Underutilization of health workers in Ghana | 100.Underutilization of health workers in Ghana | 0 | 1 | | | | | | Underutilization of health workers in Ghana | 105. Underutilization of health workers in Ghana |
| 111.106.Mandatory courses and skill sets | 101.Mandatory courses and skill sets | 0 | 1 | | | | | | Mandatory courses and skill sets | 106.Mandatory courses and skill sets |
| 112.107. Self doubt | 102.Self doubt | 0 | 1 | | | | | | Self doubt | 107. Self doubt |
| 113.108. Importance of networking and patience | 103.Importance of networking and patience | 0 | 1 | | | | | | Importance of networking and patience | 108. Importance of networking and patience |
| 114.109. Avoiding agency and debt | 104.Avoiding agency and debt | 0 | 1 | | | | | | Avoiding agency and debt | 109. Avoiding agency and debt |
| 115.110. Make the process flexible | 105.Make the process flexible | 0 | 1 | | | | | | Make the process flexible | 110. Make the process flexible |
| 116.111. Provision of mandatory training | 106.Provision of mandatory training | 0 | 1 | | | | | | Provision of mandatory training | 111. Provision of mandatory training |
| 117.112. 1.Academic Background of Researcher | 107.0 | 0 | 0 | | 1.Academic Background of Researcher | | | | 1.Academic Background of Researcher (Ai) | 112. 1.Academic Background of Researcher |
| 118.113.Confidentiality Assurance | 108.0 | 0 | 0 | | 2.Confidentiality Assurance | | | | 2.Confidentiality Assurance (Ai) | 113.Confidentiality Assurance |
| 119.114.Permission for Participation | 109.0 | 0 | 0 | | 3.Permission for Participation | | | | 3.Permission for Participation (Ai) | 114.Permission for Participation |
| 120.115.Migration Focus | 110.0 | 0 | 0 | | 4.Migration Focus | | | | 4.Migration Focus (Ai) | 115.Migration Focus |